\documentclass[aps, prb,superscriptaddress,twocolumn,showpacs,preprintnumbers,amsmath,amssymb]{revtex4}
\usepackage{amssymb}
\usepackage{graphicx}
\usepackage{dcolumn}
\usepackage{bm}
\usepackage{subfigure}
\usepackage[normalem]{ulem}

\begin{document}

\preprint{NbAs/X. J. Yang et al.}
\title{Observation of Negative Magnetoresistance and nontrivial $\pi$ Berry¡¯s phase in 3D Weyl semi-metal NbAs}

\author{Xiaojun Yang}
 \affiliation{Department of Physics and State Key Laboratory of Silicon Materials, Zhejiang University, Hangzhou 310027, China}
\author{Yupeng Li}
 \affiliation{Department of Physics and State Key Laboratory of Silicon Materials, Zhejiang University, Hangzhou 310027, China}
\author{Zhen Wang}
 \affiliation{Department of Physics and State Key Laboratory of Silicon Materials, Zhejiang University, Hangzhou 310027, China}
\author{Yi Zhen}
\email{phyzhengyi@zju.edu.cn}
 \affiliation{Department of Physics and State Key Laboratory of Silicon Materials, Zhejiang University, Hangzhou 310027, China}
 \affiliation{Collaborative Innovation Centre of Advanced Microstructures, Nanjing 210093, P. R. China}
\author{Zhu-an Xu}
 \email{zhuan@zju.edu.cn}
 \affiliation{Department of Physics and State Key Laboratory of Silicon Materials, Zhejiang University, Hangzhou 310027, China}
 \affiliation{Collaborative Innovation Centre of Advanced Microstructures, Nanjing 210093, P. R. China}

\date{\today}

\begin{abstract}
We report the electric transport properties of NbAs, which is a  Weyl semimetal candidate proposed by
recent theoretical calculations  and confirmed by recent angle-resolved photoemission spectroscopy (ARPES) data. We detected the
long-anticipated negative magneto-resistance generated by the chiral anomaly in NbAs.
Clear Shubnikov de Haas (SdH) oscillations have been detected starting from very weak
magnetic field. Analysis of the SdH peaks gives the Berry phase accumulated along the
cyclotron orbits to be $\pi$, indicating the existence of Weyl points.
\end{abstract}

\pacs{71.55.Ak, 71.70.Di, 72.15.Gd}

\maketitle

Predicted to show unprecedented features beyond the classical electronic theories of metals,
Weyl semi-metal (WSM) has motivated much interest for the realization of electronic topological properties.
The appearance of Weyl points near the Fermi level will cause
novel properties \cite{chiral_anomaly_PLB,nov_p_nc,nov_p_PRB,nov_p_PRB2,nov_p_PRB3,nov_p_PRX}. For materials with
Weyl points located near the Fermi level, called as Weyl semi-metals, exotic
low energy physics will be expected, such as the Fermi arcs on the surfaces\cite{WSM_Def,pair_PRB}, and
the chiral-anomaly induced quantum transport \cite{chiral_anomaly_PLB,CME_PRD,TaAs_arXiv_NMR}.
The band crossing points in WSMs, i.e. Weyl nodes, always
appear in pair with opposite chirality, due to the lifting of
spin degeneracy by breaking symmetry\cite{pair_PRB,Wyel SM_PRX}.
By symmetry breaking, a 3D Dirac semimetal can be driven to a
Weyl semimetal or topological insulator \cite{Cd3As2_theory_LMR}.
As a 3D analogue to graphene, the 3D Weyl semimetal could be important for
future device applications.

The chiral anomaly in WSM will in general lead to negative magneto-resistance (MR = $[\rho(H)-\rho(0)]/\rho(0)$)
when the magnetic field is parallel to the current\cite{TaAs_arXiv_NMR,Zr/HfTe5_anom_PRB}.
For ordinary metal or semiconductors the MR is weak, positive and
usually not very sensitive to the magnetic field direction.
The negative and highly anisotropic MR has been regarded as the most prominent signatures in
transport for the chiral anomaly and indicates the existence of 3D Weyl points\cite{TaAs_arXiv_NMR}.
Another distinguished feature of Dirac fermions and Weyl fermions
is the nontrivial $\pi$ Berry¡¯s phase, which results from
their cyclotron motion\cite{pi Berry theory1,pi Berry theory2}. It is a geometrical phase
factor, acquired when an electron circles a Dirac point. This
Berry¡¯s phase can be experimentally accessed by analyzing
the Shubnikov¡ªde Haas (SdH) oscillations, which has been
widely employed in 2D graphene\cite{Graphe_Berry phase1,Graphe_Berry phase2}, bulk
Rashba Semiconductor BiTeI\cite{TI_Berry phase_science}, bulk SrMnBi$_2$ in which highly anisotropic
Dirac fermions reside in the 2D Bi square net\cite{SrMnBi2_PRL},  3D Dirac
Fermions Cd$_3$As$_2$\cite{Cd3As2_Berry phase_PRL_S. Li}, and recently
in Weyl semi-metal TaAs\cite{TaAs_arXiv_Jia,TaAs_arXiv_NMR}.

Recently, using first principle calculations, Weng et al.  predicted that non-centrosymmetric
TaAs, TaP, NbAs and NbP, are time-reversal invariant 3D WSMs with a dozen
pairs of Weyl nodes which are generated by the absence of inversion center\cite{Wyel SM_PRX}.
The proposal have stimulated enormous interests. The existence of Weyl
nodes has soon been discovered in TaAs by angle-resolved photoemission spectroscopy
(ARPES) \cite{TaAs_ARPES1,TaAs_ARPES2}, and by quantum transport measurements of negative MR and a non-trivial
$\pi$ Berry¡¯s phase\cite{TaAs_arXiv_Jia,TaAs_arXiv_NMR}. Transport studies of NbAs also show ultrahigh
mobility and non-saturating MR\cite{NbAs_jpcm}, but no negative MR and a non-trivial
$\pi$ Berry¡¯s phase are acquired. However, very recent ARPES results detecte Weyl cones and Weyl
nodes in the bulk and the Fermi arcs on the surface of NbAs\cite{NbAs_ARPES}. Observation
of Negative Magnetoresistance and nontrivial $\pi$ Berry¡¯s phase in 3D Weyl semi-metal NbAs.


Niobium arsenide, NbAs, crystallizes in a body-centered tetragonal Bravais lattice, space
group I4$_1$md (109). Our X-ray diffraction (XRD) obtains lattice constants
of $a$ = 3.45 {\AA}  and $c$ = 11.68 \AA, consistent with the earlier
crystallographic studies\cite{NbAs_crystal1,NbAs_crystal2,NbAs_ARPES}.
Single crystals of NbAs were grown by vapor transport using iodine as the transport
agent, as described in Ref. \cite{NbAs_jpcm}. First, polycrystalline NbAs
was prepared by heating stoichiometric amounts of Nb and As in an evacuated silica
ampoule at 973 K for 1 day. Subsequently, the powder was loaded in a
horizontal tube furnace in which the temperature of the hot zone was kept at
1223 K  and that of the cold zone was at 1123 K. The crystals of NbAs were
verified by powder x-ray diffraction (XRD) and by compositional
analysis conducted using an energy dispersive x-ray spectroscopy (EDS). An atomic
percentage ratio of Nb:As = 49.4 : 50.6 was obtained on the EDS measurements.
The largest natural surface of the obtained NbAs single crystals was determined to be the
(112) plane by single crystal x-ray diffraction, shown in the lower inset of Fig. 1(a), with typical dimension of
1 $\times$ 1 mm$^2$. The quality of the NbAs single crystals was further
checked by the x-ray rocking curve. The full width at half maximum (FWHM) is only 0.03$^{\circ}$ (not showing here),
indicating the high quality of the single crystals. The sample was polished to a bar
shape, with 1 $\times$ 0.3 mm$^2$ in the (112) plane and a 0.2 mm thickness.
A standard six-probe method was used for both the longitudinal resistivity and transverse Hall
resistance measurements.

\begin{figure}
\includegraphics[width=8cm]{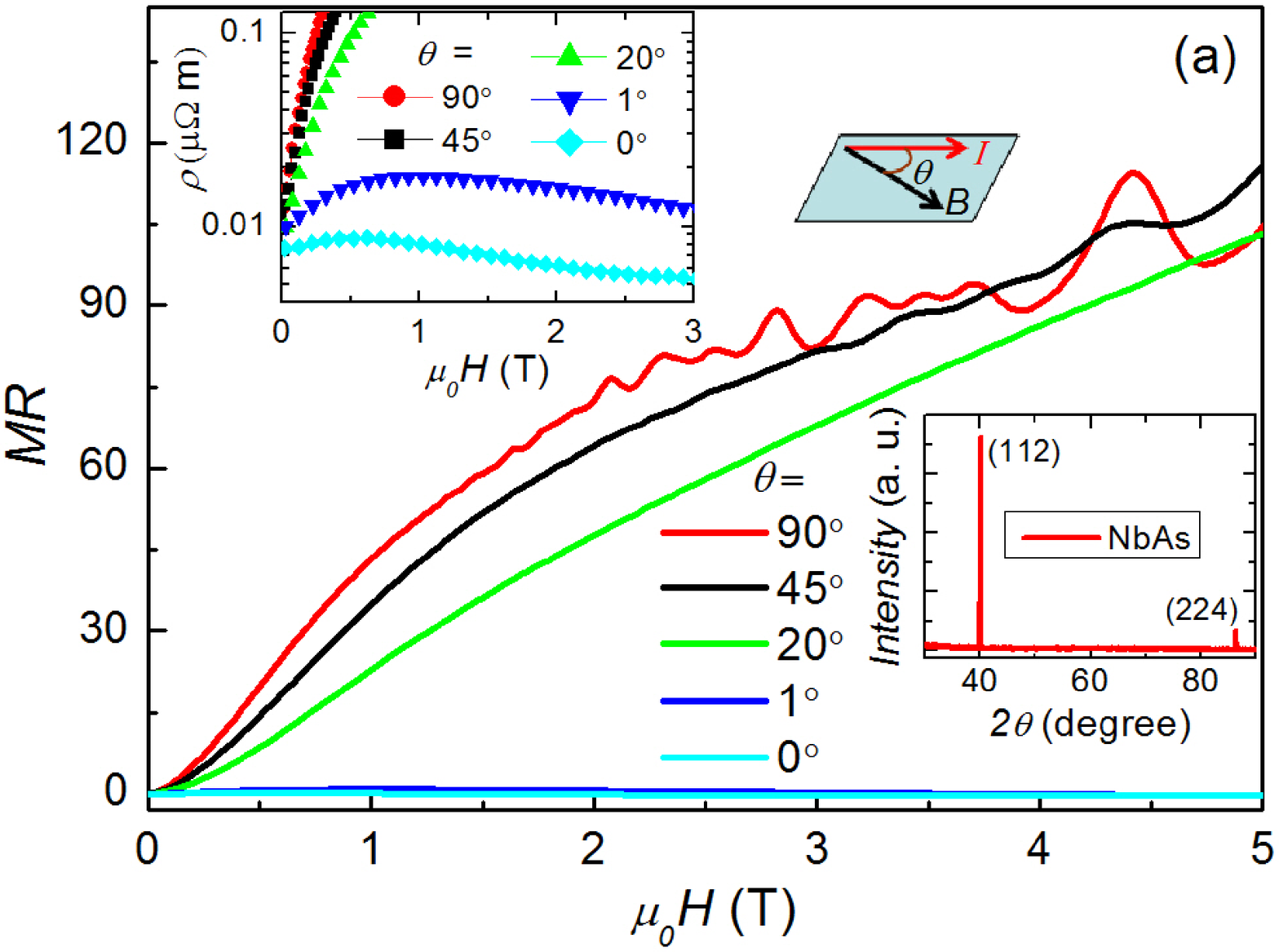}
\includegraphics[width=8cm]{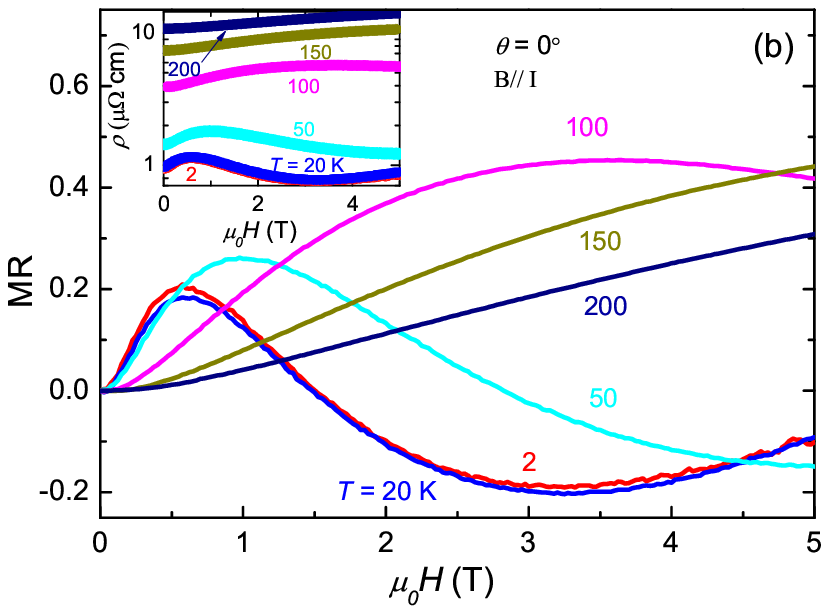}
\caption{\label{Fig.1} (Color online)  (a), Magnetic field dependence of Magnetoresistance
 with magnetic field ($\mu_0H$) from perpendicular ($\theta$ = 90$^\circ$)
 to parallel ($\theta$ = 0$^\circ$) to the electric current ($I$) at $T$ = 2 K. The upper left inset
 displays the original resistivity data plotted on logarithmic scale, emphasizing the
 contrast between extremely large positive MR for magnetic field perpendicular
 to current ($\theta$ = 90$^\circ$) and negative MR for field parallel to
 current ($\theta$ = 0$^\circ$). The upper right  inset depicts the corresponding
 measurement configurations. The lower inset present the single crystal XRD data. (b), Magnetic field dependence
 of MR under various temperatures for magnetic field ($\mu_0H$) parallel
 ($\theta$ = 0$^\circ$) to the electric current ($I$). The inset display the
 original resistivity data.}
\end{figure}

Figure 1(a) displays the field dependence of MR measured at $T$ = 2 K for several
angles $\theta$ of the applied magnetic field ($\mu_0H$) with respect to the electric
current ($I$). The angle rotates from $\mu_0H//I$ to $\mu_0H\bot I$, so that at $\theta$ = 0$^\circ$,
the applied field is parallel to the current ($\mu_0H//I$), which is the so-called
Lorentz force free configuration. When the magnetic field is applied perpendicular to the
current ($\mu_0H\bot I$, $\theta$ = 90$^\circ$), a positive MR of up to 10000\% is
observed.  This large transverse MR strongly relies on $\theta$. When the magnetic
field is rotated to be parallel to the electric current ($\theta$ = 0$^\circ$),
we get a negative MR, which is an indication of chiral magnetic field and should
be a strong evidence of Wely fermions in NbAs. The left inset of Fig. 1(a)
displays the original resistivity data plotted on logarithmic scale, emphasizing the
contrast between extremely large positive MR for magnetic field perpendicular
to current ($\theta$ = 90$^\circ$) and negative MR for field parallel to
current ($\theta$ = 0$^\circ$). The right inset of Fig. 1(a) depicts the corresponding
measurement configurations. Fig. 1(b) presents the MR at various temperatures, when magnetic field
parallel to the current($\mu_0H//I$, $\theta = 0^\circ$ ). And the inset of Fig. 1(b)
display the original resistivity data. Below 20 K, an negative MR of about -20\% can be
observed under an applied field of 3.2 T. Then, the negative MR is suppressed with
increasing temperature, and ultimately disappeared at higher temperature,
similar with the result in ZrTe$_5$\cite{ZrTe5_arXiv} and TaAs\cite{TaAs_arXiv_NMR}.

The existence of chiral quasi-particles in Dirac and Weyl semimetals opens the
possibility to observe the effects of the chiral anomaly\cite{chiral_anomaly_PLB}. The chiral
magnetic effect, which is the generation of electric current in an external magnetic field
induced by the chirality imbalance, is of  particular interest\cite{CME_PRD,CME_review}.
The most prominent signature of the chiral magnetic effect in Dirac systems in parallel electric and magnetic
fields is the positive contribution to the conductivity that has a quadratic dependence
on magnetic field\cite{CME_PRD,CME_PRB,CME_PRL}. The chiral anomaly contributed conductivity as
$\sigma = (e^3 v_F^3/4\pi^2\hbar\mu^2c)B^2$ where $\tau$ is the inter valley scattering time,
$v_F$ is the Fermi velocity near the Weyl points and $\mu$ denotes the chemical potential
measured from the energy of the Weyl points\cite{TaAs_arXiv_NMR}. This is because the  chiral
magnetic effect current is proportional to the product of
chirality imbalance and the magnetic field, and the chirality imbalance in Dirac systems is
generated dynamically through the anomaly with a rate that is proportional to the product
of electric and magnetic fields \emph{\textbf{E}}$\cdot$\emph{\textbf{B}}\cite{ZrTe5_arXiv}. As a result, the longitudinal MR becomes
negative\cite{CME_PRB,CME_PRL}, which has the maximum effect with \emph{\textbf{E}}
parallel to \emph{\textbf{B}}\cite{TaAs_arXiv_NMR}. Of course
the total conductivity of the system will also include other contributions from the nonchiral
states as well, which may weaken the negative MR effect or even overwhelm it
if the non-chiral part dominates the DC transport, which may be the case in ref.\cite{NbAs_jpcm}.
Therefore in order to see the chiral negative MR, the high quality sample with
chemical potential close enough to the Weyl point is crucial\cite{TaAs_arXiv_NMR}.

At low field, the data show sharp dips. The origin of this feature is not
completely understood, but which may be attributed to the
WAL effect stemming from the strong spin-orbit interactions\cite{WAL_PRL}, which dominates
the transport behavior of the non-chiral states\cite{ZrTe5_arXiv,TaAs_arXiv_NMR}.
When magnetic field higher than 3.2 T, the MR tend to be positive
again. This behavior is very similar to the situation in Bi$_x$Sb$_{1-x}$  and
TaAs\cite{TaAs_arXiv_NMR}. 
One possible explanation was given by X. Huang $et$ $al$. in Ref. \cite{TaAs_arXiv_NMR} The phenomena may due to the Coulomb
interaction among the electrons occupying the chiral states. Since the degeneracy of
the chiral states as well as the density of states at the Fermi level goes linearly with
the magnetic field, eventually the system will approach a spin-density-wave (SDW)
like instability under Coulomb interaction\cite{NTPMR_PRB}. Then at finite temperature, the strong
SDW fluctuation provides another scattering channel which can be greatly enhanced
in high field and may give the positive MR in the high field region\cite{TaAs_arXiv_NMR}.

\begin{figure}
\includegraphics[width = 8cm]{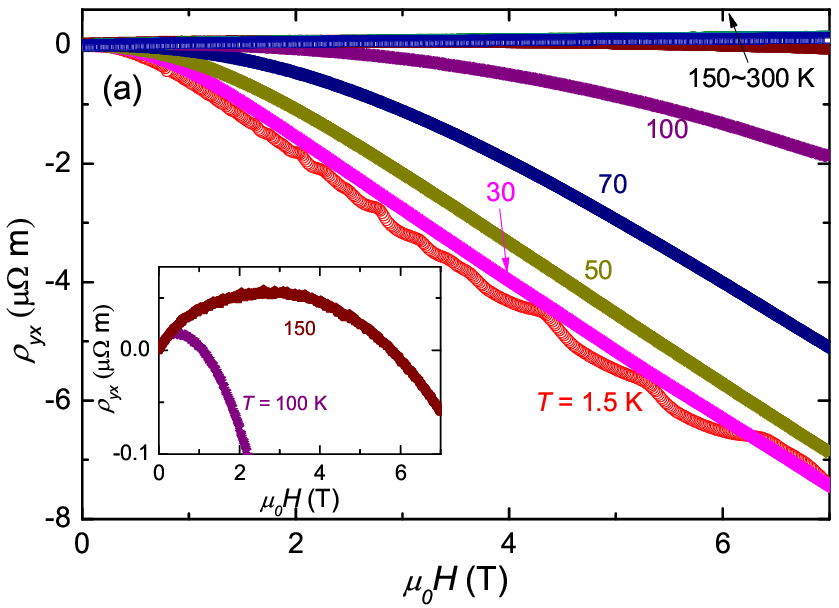}
\includegraphics[width = 8cm]{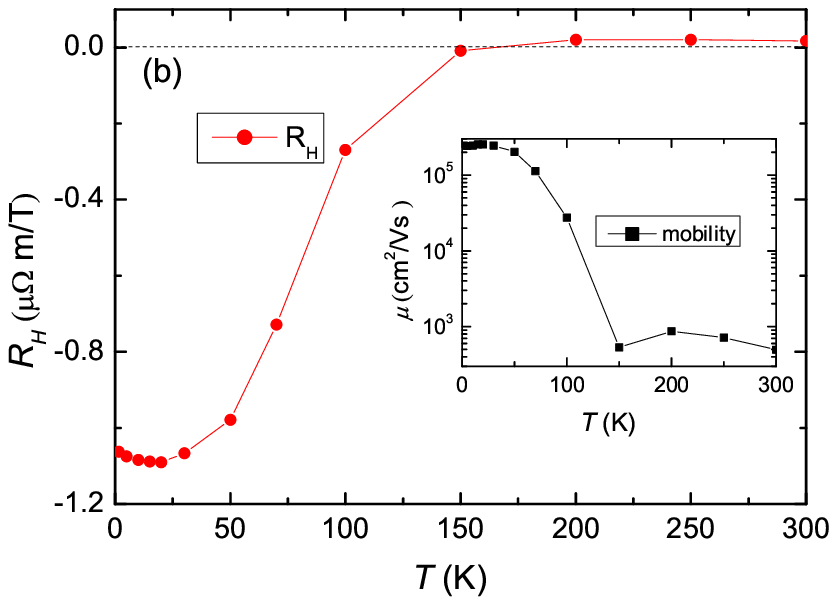}
\caption{\label{Fig.2}(Color online) (a), Hall resistivity measured at various
temperatures from 2 to 300 K. The inset displays enlarged plot of
Hall resistivity curves at $T$ = 100 K and 150 K. (b), Hall coefficient $R_H$ at 7 T
as a function of temperature. The inset displays mobility versus
temperature determined by the Hall coefficient at 7 T and the zero field resistivity
using a single band approximation.}
\end{figure}

Figure 2(a) displays the magnetic field dependence of Hall resistivity $\rho_{yx}$(B//c) measured at
various temperatures. At low temperature, the negative slope in high magnetic fields
indicates that the electrons dominate the main transport processes. However, in low
fields the curve tends to be flat. At 100 K and 150 K, for example, $\rho_{yx}$ is initially positive under low
fields but changes to negative in higher fields.
The curvature and sign reversal of the Hall resistivity indicates
the coexistence of hole-type minority carriers with high mobility
and electron-type majority carriers with low mobility.
At higher temperature slope of Hall
resistivity changes to positive, implying the carriers dominating
the conduction mechanism transformed to hole-type. All these are
consistent with multiple hole- and electron-like carriers
as was also observed in TaAs and NbP, and indicated by band
structure calculations\cite{NbP_arXiv,TaAs_arXiv_Jia,TaAs_arXiv_NMR,Wyel SM_PRX,two band_theory_arxiv1}
and a previous paper on NbAs\cite{NbAs_jpcm}. As shown in  Fig. 2(b), the
material show negative Hall coefficient, $R_H(T)$, up to 150 K and then changes sign for
temperature above about 150 K. For the sake of simplicity, we use the single  band
to estimate the mobility. The inset of Fig 2(b) displays mobility versus
temperature determined by the Hall coefficient at 7 T and the zero field resistivity
using a single band approximation. The mobility plays a
major role for charge transport in a material, and consequently decides the efficiency of
various devices. Here, NbAs exhibits an ultrahigh mobility of
2.45 $\times$ 10$^5$ cm$^2$/Vs at 1.5 K, consistent with the result in ref. \cite{NbAs_jpcm}.

\begin{figure}
\includegraphics[width=8cm]{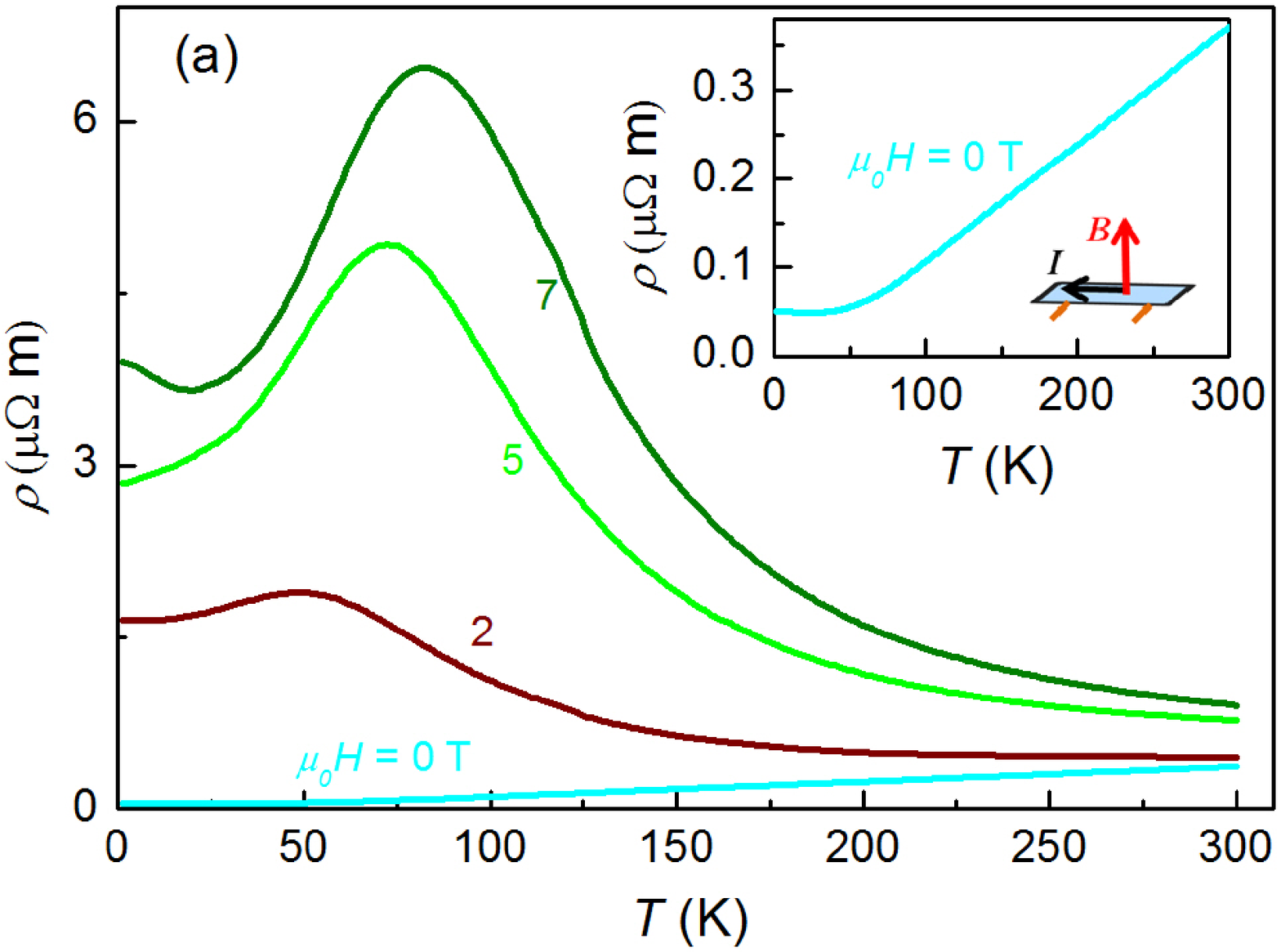}
\includegraphics[width=8cm]{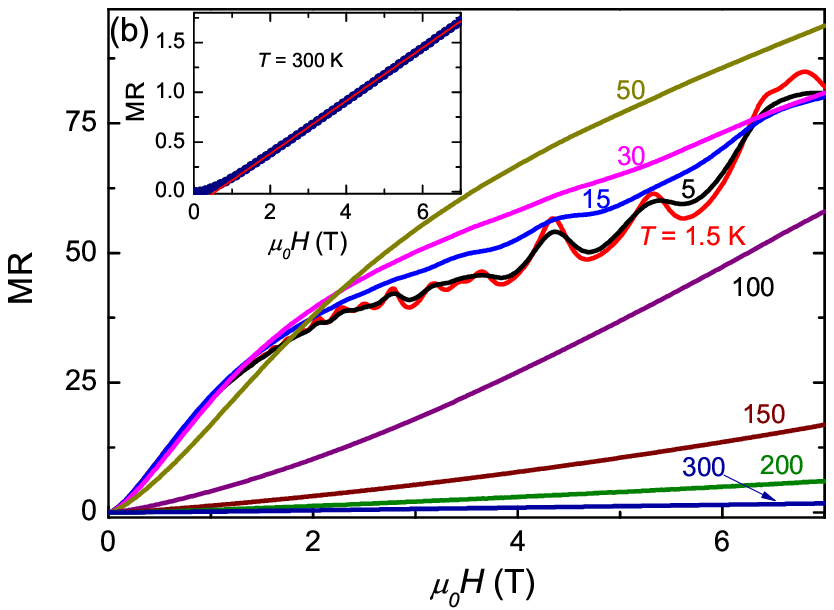}
\caption{\label{Fig.3}(Color online)  (a), The temperature dependence of
resistivity in magnetic field perpendicular to the electrical current.
The inset of (a) gives the measurement configuration, and zooms in on the case of 0 T.
(b) Magnetic field dependence of MR at representative temperatures. The inset
of (b) displays the enlarged plot for $T$ = 300 K.}
\end{figure}

In Fig. 3(a) the temperature dependence of resistivity  is plotted. The inset
gives the measurement configuration. In zero
magnetic field, NbAs exhibits a metallic behavior down to 1.5 K.
The applied magnetic field not only significantly increases the resistivity, but also
stimulates a crossover from metallic to insulator like behavior, which may be related
to the formation of the Landau levels under magnetic field\cite{TaAs_arXiv_NMR}.
Under applied field, a resistivity anomaly can be observed,  which can be enhanced
and move to high temperature under higher magnetic fields. Similar behavior was
observed in TaAs\cite{TaAs_arXiv_NMR}, ZrTe$_5$ and HfTe$_5$\cite{Zr/HfTe5_anom_PRB}.

Figure 3(b) displays the field dependence of MR at various temperatures
At 300 K, the MR reaches as high as 170\% at 7 T, as shown in the
inset of Fig. 3(b). Above 1.5 T, the MR of our
NbAs single crystal is quite linear, which is consistent
with the theoretical prediction of  linear quantum MR even up to the
room temperature induced by 3D Dirac cone type electronic
structure\cite{Cd3As2_theory_LMR}. However, the quantum
limit required in Ref.\cite{Cd3As2_theory_LMR} is actually not reached in our
sample, since there is clearly more than one Landau level
occupied in our field range, as will be seen in Fig. 4.
Similar situations exist in multilayer epitaxial graphene\cite{Graphe_LMR_NL},
the topological insulators Bi$_2$Se$_3$ and Bi$_2$Te$_3$\cite{BiSe_LMR_AN,BiSe_LMR_PRL,BiSe_LMR_Science}
 and Dirac Fermions Cd$_3$As$_2$\cite{Cd3As2_Berry phase_PRL_S. Li}. To understand the physical origin of the linear
MR without reaching the quantum limit, further theoretical
studies are highly desired. Nevertheless, this large room temperature
linear MR is quite unusual. If its magnitude
can be further enhanced, NbAs may be useful for
practical applications in magnetic random access memory
and magnetic sensors. At low temperatures, clear Shubnikov de Haas (SdH)
oscillations have been detected starting from very weak
magnetic field.

\begin{figure}
\includegraphics[width=8cm]{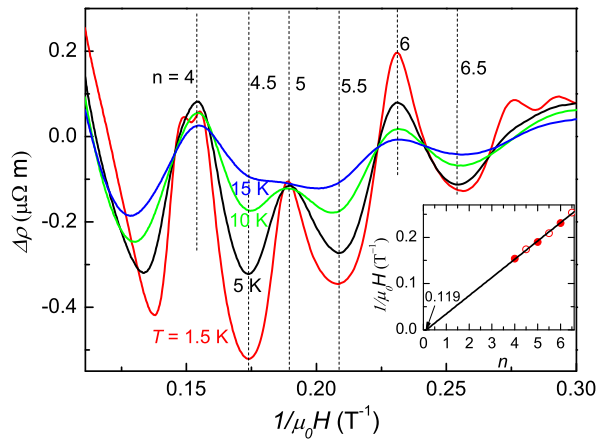}
\caption{\label{Fig.4}(Color online)  The high field
oscillatory component $\vartriangle\rho$ plotted against inverse field 1/$\mu_0H$ at
$T$ = 1.5, 5, 10 and 15 K.  The inset displays SdH fan
diagram plotting the measured1/$B_n$ with the filling factor n, which
is estimated from the $\vartriangle\rho$ versus 1/$B_n$ plot.  }
\end{figure}

Figure 4 shows the oscillatory
component of $\Delta\rho$ versus $1/B$ at various temperatures
after subtracting a smooth background. They are periodic
in $1/B$, as expected from the successive emptying of the
Landau levels when the magnetic field is increased.
The peaks marked by Landau index n are mainly stemming
from the oscillations with the frequency of 25 T.  The
cross-sectional area of the Fermi surface (FS) $A_F = 2.38 \times 10^{-3} \text{\AA}^{-2}$ can be obtained
according to the Onsager relation $F = (\Phi_0/2\pi^2)/A_F$, where $F$ is oscillation frequency
and $\Phi_0$ is flux quantum. We assign integer indices to
the $\Delta\rho$ peak positions in $1/B$ and half integer indices to
the $\Delta\rho$ valley positions. According to the Lifshitz-Onsager
quantization rule $A_F(\hbar/eB) = 2\pi(n+1/2+\beta+\delta)$, the
Landau index $n$ is linearly dependent on $1/B$. 2$\pi\beta$ is the
Berry's phase, and 2$\pi\delta$ is an additional phase shift resulting
from the curvature of the Fermi surface in the third
direction\cite{delta_science}. $\delta$ changes from 0 for a quasi-2D cylindrical
Fermi surface to $\pm$1/8 for a corrugated 3D Fermi surface\cite{delta_science,Cd3As2_Berry phase_PRL_S. Li}.
Our data points in Fig. 4 can be fitted linearly, and the linear
extrapolation gives an intercept 0.119, which means a non-trival $\pi$
Berry's phase ($\beta = 1/2$) and $\delta = 0.119$ (very close to 1/8),
signaling the 3D Dirac fermion behaviors, and implying the
FS associated with 25 T quantum oscillations is enclosing a Weyl point.

In the trivial parabolic dispersion case such as that
involving conventional metals, the Berry's phase 2$\pi\beta$
should be zero. For Dirac systems with linear dispersion,
there should be a nontrivial $\pi$ Berry's phase ($\beta$ = 1/2). This
non-trivial $\pi$ Berry's phase has been clearly observed in 2D
graphene\cite{Graphe_Berry phase1,Graphe_Berry phase2}, bulk
Rashba Semiconductor BiTeI\cite{TI_Berry phase_science}, bulk SrMnBi$_2$ in which highly anisotropic
Dirac fermions reside in the 2D Bi square net\cite{SrMnBi2_PRL}, 3D Dirac
Fermions Cd$_3$As$_2$\cite{Cd3As2_Berry phase_PRL_S. Li}, and recently
in Weyl semi-metal TaAs\cite{TaAs_arXiv_Jia,TaAs_arXiv_NMR}.
The intercept 0.119 obtained in Fig. 4 clearly
reveals the nontrivial $\pi$ Berry's phase, and thus provides strong
evidence for the existence of Weyl fermions in NbAs.
An additional phase shift $\delta$ = 0.119 $\sim$ 1/8 should result from the 3D nature
of the system. Complementary to previous theoretic\cite{Wyel SM_PRX} and ARPES experiments
result\cite{NbAs_ARPES}, our bulk transport measurements
confirm the 3D Dirac semimetal phase in NbAs.

In summary, we have performed bulk transport measurements on single crystals
of the proposed 3D Weyl semimetal NbAs.  Large MR as high as 10000\% is detected
with magnetic field perpendicular to the current. When the external magnetic field
is rotated to be parallel with the current, Chiral anomaly induced negative MR up
to -20\% is observed. This unusual negative MR is the first electric transport evidence for the
chiral anomaly associated with the Weyl points in NbAs. Hall effect data
suggest that both n and p types of carriers exist in NbAs.
NbAs exhibits an ultrahigh mobility of 2.45 $\times$ 10$^5$ cm$^2$/Vs at 1.5 K.
A large linear quantum magnetoresistance is observed at room temperature.
By analyzing the Shubnikov¡ªde Haas oscillations of longitudinal resistance
at low temperature, a nontrivial $\pi$ Berry¡¯s phase with a
phase shift of 0.119 is obtained, which provides bulk quantum
transport evidence for the existence of a 3D Wyel semimetal
phase in NbAs. With its unique electronic structure,
unusual high mobility, and large room-temperature linear
magnetoresistance, the 3D Wyel semimetal NbAs may
open new avenues for future device applications.

This work is supported  by the National Basic Research Program of
China (Grant Nos. 2011CBA00103 and 2012CB821404), NSF of China
(Contract Nos. 11174247 and U1332209), Specialized Research Fund
for the Doctoral Program of Higher Education (Grant No.
20100101110004), and the Zhejiang Provincial Natural Science
Foundation of China (Grant. No. Y6100216).

\end{document}